\documentclass[10pt,conference]{IEEEtran}
\usepackage{pgfplots}
\pgfplotsset{compat=1.18}
\IEEEoverridecommandlockouts
\usepackage{cite}
\usepackage{amsmath,amssymb,amsfonts}
\usepackage{algorithmic}
\usepackage{graphicx}
\usepackage{svg}
\usepackage{lipsum} 
\usepackage{textcomp}

\usepackage[english]{babel}
\usepackage{lipsum}
\usepackage{soul}
\usepackage{multirow}
\PassOptionsToPackage{table,xcdraw}{xcolor}\usepackage{xcolor}
\usepackage{tikz}
\usepackage{mdframed}

\usepackage{pgfplots}

\definecolor{mygray}{RGB}{211,211,211} 

\def\BibTeX{{\rm B\kern-.05em{\sc i\kern-.025em b}\kern-.08em
    T\kern-.1667em\lower.7ex\hbox{E}\kern-.125emX}}
\begin{document}

\title{Embracing Experiential Learning: Hackathons as an Educational Strategy for Shaping Soft Skills in Software Engineering
\thanks{This study was financed in part by the Coordenação de Aperfeiçoamento de Nível Superior - Brasil (CAPES) - Postdoctoral Institutional Program (PIPD).}
}

\author{\IEEEauthorblockN{Allysson Allex Araújo}
\IEEEauthorblockA{\textit{Center for Science and Technology} \\
\textit{Federal University of Cariri}\\
Juazeiro do Norte, Brazil \\
allysson.araujo@ufca.edu.br}
\and
\IEEEauthorblockN{Marcos Kalinowski}
\IEEEauthorblockA{\textit{Department of Informatics} \\
\textit{Pontifical Catholic University of Rio de Janeiro}\\
Rio de Janeiro, Brazil \\
kalinowski@inf.puc-rio.br}
\and
\IEEEauthorblockN{Maria Teresa Baldassarre}
\IEEEauthorblockA{\textit{Department of Computer Science} \\
\textit{University of Bari}\\
Bari, Italy \\
mariateresa.baldassarre@uniba.it}
}

\maketitle

\begin{abstract}
In recent years, Software Engineering (SE) scholars and practitioners have emphasized the importance of integrating soft skills into SE education. However, teaching and learning soft skills are complex, as they cannot be acquired passively through raw knowledge acquisition. On the other hand, hackathons have attracted increasing attention due to their experiential, collaborative, and intensive nature, which certain tasks could be similar to real-world software development. This paper aims to discuss the idea of hackathons as an educational strategy for shaping SE students' soft skills in practice. Initially, we overview the existing literature on soft skills and hackathons in SE education. Then, we report preliminary empirical evidence from a seven-day hybrid hackathon involving 40 students. We assess how the hackathon experience promoted innovative and creative thinking, collaboration and teamwork, and knowledge application among participants through a structured questionnaire designed to evaluate students' self-awareness. Lastly, our findings and new directions are a\-nalyzed through the lens of Self-Determination Theory, which offers a  psychological lens to understand human behavior. This paper contributes to academia by advocating the potential of hackathons in SE education and proposing concrete plans for future research within SDT. For industry, our discussion has implications around developing soft skills in future SE professionals, thereby enhancing their employability and readiness in the software market.
\end{abstract}


\section{Introduction}



Alongside the last years, SE scholars \cite{capretz2018call, akdur2022analysis} and practitioners have been arguing about the relevance of soft skills in the practice of SE and the need to incorporate this topic in the context of SE education. As discussed by Sedelmaier and Landes \cite{sedelmaier2014practicing}, soft skills are core competencies of a software engineer and for this reason soft skills should be a core part of SE education. However, one of the difficulties is in selecting the right teaching or training approach that allows a person to gain new abilities in an environment that is somewhat similar to the nuances in the market \cite{matturro2019systematic}. In this sense, it became evident that the SE community has been working to figure out how to improve the soft skills of SE students. For instance, Dubey and Tewari \cite{dubey2016systematic}, reviewed SE curricula and found references to soft skills such as communication and leadership. Marques et al. \cite{marques2020stimulating} explored how Design Thinking provides practical experiences and it supports the development of students' soft skills from SE courses. Garcia et al. \cite{garcia2020effects}, in turn, investigated the effects of game-based learning in the acquisition of soft skills on undergraduate SE courses.

As one can notice, these pedagogical approaches to teaching soft skills exhibit a high degree of variability, as it is commonly accepted that soft skills cannot be learned passively through raw knowledge acquisition \cite{martins2007role}. Caeiro-Rodríguez et al. \cite{caeiro2021teaching}, for example, clarified that students must take an active role, experiencing their capabilities, strengths, and weaknesses in relation to soft skills. Given this scenario, one can highlight the potential of hackathons in leveraging soft skills for SE students by bridging theory to practice for analogous real-world nuances (time pressure, collaborative teamwork, innovative thinking, etc.) \cite{porras2018hackathons}. In summary, a hackathon is an intensive and collaborative event where participants, typically software developers, engage in focused, time-limited programming and problem-solving activities \cite{steglich2021online, sadovykh2020hackathons}. Hackathons usually range from a few hours to a few days and are often organized around a theme/challenge where the participants can work in teams to develop and present software solutions \cite{steglich2020hackathons}. These events are known for their fast-paced and dynamic nature, encouraging creativity and teamwork, as participants strive to create functional software within a condensed timeframe \cite{kumalakov2018hackathon}. 

Indeed, the prominence of hackathons as immersive and experiential learning experiences able to enhance students' skill sets has already garnered considerable interest among SE researchers \cite{steglich2020hackathons, steglich2021online, sadovykh2020hackathons, gama2018hackathon, kumalakov2018hackathon, afshar2022hackathon}. In particular, our paper specifically aims to discuss the idea of hackathons in enhancing soft skills of SE students. We believe that this effort can help raise awareness to this educational opportunity and its importance for both academia and industry. To achieve this objective, we have defined the following research question: ``\textit{Can participation in hackathons lead to a self-perception of improvement in soft skills for SE students?}''. To investigate this issue, we first briefly overview the literature on soft skills and hackathons in SE education. Additionally, we present a case study that brings initial findings from a hackathon conducted at a public university in Brazil. We also provide details on how we organized our hackathon in an educational setting, including operational details and the lessons we learned. Lastly, we discuss our findings in the light of Self-determination Theory (SDT) \cite{deci2012self}, which served as a psychological lens for deriving relevant implications for future research and practice.

Our investigated hackathon spanned seven days and blended online and in-person activities. It attracted 40 enrolled students, 79\% of whom were participating in a hackathon for the first time. This research reports a preliminary assessment with a formal evaluation strategy covering how engagement in the hackathon fostered creativity, innovative thinking, collaboration, teamwork, and knowledge application among students. These specific soft skills were investigated due to their critical relevance in enhancing overall skill development, particularly in enhancing the synergy between software development and business \cite{moreira2023analyzing}. Naturally, the empirical findings we report in this work are initial evidence, emphasizing the need for more investigation to understand the intricacies of this challenge. 

Furthermore, this paper is part of an ongoing project that focused on behavioral aspects related to hackathon. In contrast to this previous work \cite{araujo2024can}, our primary goal with this paper is to raise awareness and discuss a forward-thinking idea about the potential value of hackathons in helping SE students bridge the gap between technical proficiency and practical soft skills grounded on the experiential learning and SDT. We did not address the perspective of soft skills in our previous paper.

This study offers three main contributions. For \textit{academia}, it discusses the potential of hackathons to enhance soft skills in SE students, supported by preliminary empirical evidence. For \textit{practice/industry}, it highlights the role of hackathons in developing soft skills in SE students, enhancing their employability and readiness for the  software market. As a \textit{forward-looking idea}, it suggests directions based on SDT to explore hackathons to develop soft skills in SE students, thus contributing to ongoing advancements in SE practice.


\section{Background and Related Work}


\subsubsection{Unveiling the Challenge: Enhancing Soft Skills in SE Students}

Traditionally, SE programs have primarily emphasized technical skills (also known as ``hard skills'') like programming, software testing, and system architecture. However, the constantly evolving nature of the software industry and the collaborative demands of modern workplaces have emphasized the importance of soft skills. These skills encompass a broad range of characteristics, including personality traits, social interaction abilities, communication, and personal habits \cite{ahmed2012evaluating}. According to Kechagias \cite{kechagias2011teaching}, soft skills can be defined as ``intra and inter personal (socio-emotional) skills'' that are essential for personal development, social participation, and workplace success. Capretz and Ahmed \cite{capretz2018call} advocate for promoting soft skills in SE to enhance career prospects and better prepare practitioners for the industry's challenges. This need is also driven by the increasing complexity and interdisciplinary nature of software projects \cite{matturro2019systematic, miranda2021compreendendo}.

From an educational lens, teaching soft skills presents challenges because these skills cannot be passively acquired through theoretical knowledge alone \cite{martins2007role}. SE scholars have been investigated active learning methods to enhance students' soft skills. These proposals include, for example, project-based learning \cite{gonzalez2011teaching}, game-based learning \cite{garcia2020effects}, and problem-based learning \cite{yu2016developing}.  While these approaches are valuable and contribute to soft skill development, they may not fully replicate the fast-paced, high-pressure, and collaborative nature of software projects. Hackathons, in contrast, provide an immersive environment where students must rapidly adapt and collaborate under specific constraints, making them a promising complementary strategy for enhancing soft skills in SE education \cite{porras2018hackathons, steglich2021online}.

\subsubsection{Hackathons in SE Education: Leveraging Experiential Learning}
The importance of developing soft skills in SE students has also been successfully linked to Experiential Learning (ExL) \cite{fioravanti2023software}. ExL is based on learning through direct experience, specifically defined as ``learning through reflection on doing'' \cite{bradberry2019learning}. By involving students in hands-on activities and encouraging reflection, ExL helps bridge the gap between theoretical concepts taught in the classroom and real-world scenarios. This educational approach resembles the essence of hackathons, where participants engage in intense problem-solving sessions and later reflect on their experiences.

Researchers have been closely studying the pedagogical potential of hackathons in SE education, focusing on various aspects such as skill development, student perceptions, and curriculum integration. This focus on hackathons stems from their ability to provide a rich experiential learning approach where students can apply knowledge in practice. Porras et al. \cite{porras2018hackathons} conducted a literature review, utilizing their extensive learning experiences and written and interview materials from students and industry participants to present an overview of hackathons. In another study, Porras et al. \cite{porras2019code} discussed a systematic literature review on the use of hackathons in SE and computer science education.


Other works have focused on the impact of hackathons on student skills and perceptions.  Steglich et al. \cite{steglich2020hackathons} conducted a case study to investigate factors influencing student participation, their perceptions, and the SE practices they adopted. Steglich et al. \cite{steglich2021online} further examined the professional skills developed during hackathons and how intense collaboration occurs in atypical scenarios. In addition, Haque et al. \cite{haque2022effectiveness} assessed students' ability to apply SE best practices learned in their courses to hackathon settings, demonstrating the practical application of these skills.

Hackathons have also been integrated into SE curricula.  Sadovykh et al. \cite{sadovykh2020hackathons} introduced hackathons as part of an SE program, proposing a specific setup and evaluating outcomes through surveys from students, mentors, and industry representatives. They reported high satisfaction and suitability for final evaluations in Software Quality courses. Sadovykh et al. \cite{sadovykh4hackathon} also found that hackathons as exam replacements motivated students to learn beyond the course material. Gama et al. \cite{gama2018hackathon} proposed a methodology for project-oriented undergraduate courses, using hackathons as realistic scenarios to practice class knowledge, and evaluated its use in an IoT course through quantitative and qualitative analyses. Kumalakov et al. \cite{kumalakov2018hackathon} explored a blended hackathon model to increase student engagement and interest in computer science, with preliminary findings indicating positive results.

In terms of practical applications, Afshar et al. \cite{afshar2022hackathon} integrated a hackathon component into a software development and architecture course. Data collected from pre- and post-hackathon surveys and GitHub code commits demonstrated the impact of hackathon participation on student performance and understanding. La Place and Jordan \cite{la2022adapting} found that students who participated in hackathons and a project-based learning (PBL) SE degree developed transferable skills and employed various problem-solving approaches in both settings. Unlike traditional PBL, which focuses on long-term collaboration and gradual skill development, hackathons prioritize immediate problem-solving under time constraints \cite{szymanska2020effects}. This high-pressure setting inherent to hackathons enhances soft skills like communication, leadership, and conflict resolution, complementing the gradual learning process in PBL \cite{garcia2023fostering}.

\subsubsection{Research Gap} Although some papers have previously addressed this issue, there is a scarcity of studies specifically focusing on the role of hackathons in developing soft skills among SE students. We found no other work discussing these results through the lens of SDT. Further exploration of these aspects could help optimize hackathon design to enhance their educational impact and bridge the gap between technical proficiency and soft skills development in SE education.

\section{Method}

\subsection{Research Design}
This investigation adopts an evaluative case study approach following the methodological guidelines proposed by Runeson and Höst \cite{runeson2009guidelines}. Our investigation seeks to tackle the following \textbf{research question} (1): \textit{Can participation in hackathons lead to a self-perception of improvement in soft skills for SE students?} As we can see, this research question  aligns with the open call concerning the relevance of approaching soft skills into SE education \cite{akdur2022analysis, sedelmaier2014practicing}.

In the \textbf{case selection} (2), students from a public university in Brazil participated in a sponsored educational hackathon. This case was selected for its relevance to the study’s aim of exploring hackathons’ role in developing SE students’ soft skills. The sample included bachelor’s students who expressed interest, with selection based on event enrollment.



Concerning \textbf{data collection} (3), all the participants had to initially answer a pre-event questionnaire designed to collect demographic information like age, gender, course of study, and the duration of their enrollment at the university. 
Following the hackathon's conclusion, participants completed a post-event questionnaire featuring scaled questions (from 1 to 7) designed to evaluate the extent of improvement in three key perspectives: Q1) How much did participation in the hackathon stimulate your creativity and innovation?	Q2) How much did the hackathon provide a collaborative and teamwork environment? Q3) How much did the hackathon allow you to apply the knowledge acquired in the classroom? 




The \textbf{data analysis} (4) encompassed quantitative data from questionnaires responses, capturing demographic details and scaled questions related to soft skills. Descriptive statistical analysis was performed utilizing measures such as means, standard deviations, and frequency distributions. Lastly, we discuss our findings in the light of Self Determination Theory, connecting our empirical results with a theoretical background.




To increase \textbf{validity of the findings} (5), we followed the well-known case study protocol recommended by Runeson and Höst \cite{runeson2009guidelines}. To enhance the verifiability and transparency of our results, we have also made all our data openly available through our supporting repository \cite{repo}. All participants were  informed about the study's context, including its objective, required data, potential risks, and other relevant details, with explicit consent obtained through consent form.


\subsection{Case Study Description and Participants Characterization}
Ten teams of four students participated in a seven-day hackathon from May 18 to May 24, 2023. Teams were selected based on their registration order, with the final list announced on May 15, 2023. The hybrid format required strong self-management skills for both online and in-person activities.

Concerning the online activities, we set up a Discord server accessible to all event participants, including team members, mentors, and organizers. The server was organized into three categories, namely 1) Workspace, 2) Information, and 3) Organization. The \textit{Workspace} category included one text channel and one voice channel for each team. These channels were only accessible by the organizers and respective team members. In addition, the \textit{Information} comprised channels that were open to all participants, with specific channels dedicated to important information such as welcome messages, challenge details, sponsor advertise, off-topic discussions, and support. The \textit{Organization} included exclusive channels for members of the organizing committee and mentors. These channels were useful to follow up and manage communication.

In regard to in-person activities, teams were granted access to the university's laboratory facilities based on their specific demands. These demands were gathered through the enrollment process, during which participants registered their teams and specified whether they would need access to the laboratory and at what time. Each team was assigned mentors who were alumni of the university and currently employed in the software industry. This partnership with alumni was a strategic decision due to two primary reasons. First, it facilitated the connection between current students and recent graduates, fostering a culture of knowledge exchange and networking. Second, it evidenced a sense of gratitude from mentors who appreciated the opportunity to return to the university and contribute back to their community.

The problem addressed during the hackathon was the development of a meeting tracking software for undergraduate thesis supervision. A software requirements document was used to inform the participants about the product requirements specification. On the last day of the hackathon, each team had to present a minimally functioning prototype that included both front-end and back-end elements. The presentation took the form of a structured slide deck pitch. Only one team opted not to present their pitch for personal reasons. The solutions were evaluated by a panel of five invited judges, who assessed criteria such as creativity and innovation, quality of design and usability, functionality and feasibility, and presentation (oral and visual). The team with the highest average score was declared the winner. Members of the winning team were awarded personalized small 3D trophies crafted using a printer from the own university, along with sponsored gift vouchers as symbolic prizes. Additionally, all teams that presented their projects received a certificate of completion.

Out of 40 participants who enrolled, we obtained 34 responses (85\%) for the pre-hackathon questionnaire and 35 responses (87.5\%) for the post-hackathon questionnaire. 
Among the participants, the vast majority (over 80\%) were male, based on responses to the gender question, which offered the options: male, female, non-binary, prefer not to answer, and an open choice (to be filled out by the participant). Most of the respondents (79.40\%) fell within the age range of 18-21, whereas 14.70\% were aged between 22-25, and 2.90\% each were aged between 26 and older. Regarding the distribution of courses, 50\% of the participants were from the Information Systems course, 47.10\% were from Computer Science, and 2.90\% were from Mining Engineering. In terms of enrollment duration at university, 26.50\% of students had been enrolled for less than a year, 61.80\% for one to three years, and the remaining 11.80\% for over three years.

\section{Preliminary Empirical Findings}

\begin{figure}[ht!]
  \centering
  \includegraphics[scale=1.07]{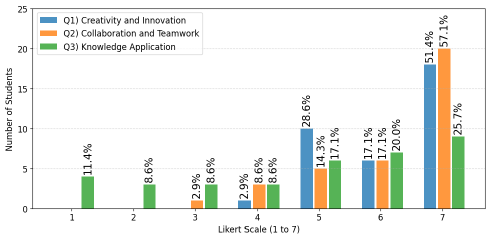}
      \caption{Distribution of responses for each question.}
  \label{fig:results}
\end{figure}

Figure \ref{fig:results} shows the distribution of responses for each question analyzed. The questions (Q1, Q2, and Q3) revolve around the innovative and creative thinking (blue), collaboration and teamwork (orange), and practical knowledge application (green). Looking at the responses for \textbf{innovative and creative thinking}, the answers ranged from 4 to 7. The majority of responses (85.7\%) were clustered around the higher end of the scale (rated 6 or 7), indicating a positive self-perception among students regarding the impact of the hackathon on their creativity and innovation. The average rating of approximately 6.12 suggests a strong overall level of stimulation experienced by the participants. The standard deviation of approximately 0.95 indicates that the responses were relatively tightly clustered around the mean, implying a consistent perception among the participants regarding the effectiveness of the hackathon in stimulating creativity and innovation. However, there were a few lower ratings, suggesting variability in individual perceptions. Of course, further analysis could explore factors influencing these perceptions, such as team dynamics, project complexity, or personal creative inclinations.

Moving on to the responses for \textbf{collaboration and teamwork}, the answers ranged from 3 to 7. The majority of responses (86.1\%) were clustered towards the higher end of the scale (rated 6 or 7), indicating a strong self-perception among participants regarding the hackathon's ability to enact collaboration and teamwork. The average rating of 6.29 further supports this positive perception, suggesting a higher level of collaboration experienced during the hackathon. The standard deviation, calculated to be approximately 1.36, indicates some variability in participants' experiences. Further analysis could explore factors influencing these perceptions, such as team dynamics, communication strategies, or task allocation, providing potential insights into the effectiveness of hackathons in fostering collaborative environments for SE students.

Finally, the responses for \textbf{practical knowledge application} ranged from 1 to 7. Notably, 47.1\% of responses fell into the lower end of the scale (rated 1 to 3), indicating a mixed perception regarding the extent to which the hackathon facilitated the application of classroom-acquired knowledge. The average rating of approximately 4.97 suggests a moderate level of success in applying classroom knowledge during the hackathon. However, the standard deviation, calculated to be approximately 2.03, indicates a certain variability in participants' experiences regarding knowledge application. One  reason for this variation is that 26\% of the participants were first-year students with limited experience in course subjects. Further analysis could investigate the correlation between classroom instruction and hackathon tasks.





\vspace{-0.4cm}
\section{Looking to the Future}

The research question behind this study is: ``\textit{Can participation in hackathons lead to a self-perception of improvement in soft skills for SE students?}''. To answer this question, we provide here an analysis covering what we found in our preliminary case study backed by what the literature has been discussing about this subject. We opted to anchor this analysis within the Self-Determination Theory (SDT) framework \cite{deci2012self} due to its well-known success in depicting motivation as a driver for sustained engagement and learning. 







\begin{figure}[ht!]
  \centering
  \includegraphics[scale=1]{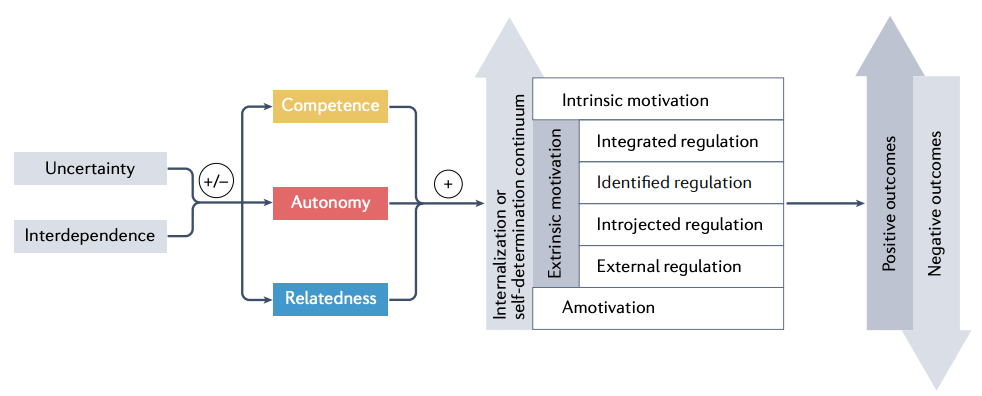}
      \caption{Primary concepts of Self-determination Theory \cite{gagne2022understanding}}
  \label{fig:sdt}
\end{figure}

In summary, SDT offers a psychological lens through which to examine human motivation and behavior, distinguished into \textit{intrinsic motivation} (doing something for its own sake, out of interest and enjoyment), \textit{extrinsic motivation} (doing something for an instrumental reason), and \textit{amotivation} (lacking any reason to engage in an activity) \cite{ryan2023self, ryan2022self}. According to Deci and Ryan \cite{deci2012self}, SDT posits that individuals are driven by three psychological needs: autonomy, competence, and relatedness. These needs impact motivation, which influences outcomes (in our research, soft skills improvement). In addition, both these needs and motivation might be influenced by the uncertainty and interdependence issues. Inspired by the work of Gagné et al. \cite{gagne2022understanding}, we approach below the SDT concepts previously explained (depicted in Figure \ref{fig:sdt}) in order to understand how hackathons may shape the soft skills of SE students.

Hackathons have been found to impact student motivation \cite{araujo2024can, steglich2020hackathons}, which may be influenced by factors such as \textbf{uncertainty} and \textbf{interdependence} \cite{ryan2022self}. As discussed by Griffin et al. \cite{griffin2007new}, higher levels of uncertainty require adaptive behaviors, while interdependence levels necessitate social, team-oriented, and network-oriented behaviors. Based on this comprehension, we may assume that the  dynamic nature of hackathons provides an ideal environment for students to develop and practice their soft skills, including how to handle uncertainty and interdependence effectively with their teams. Our preliminary results, for example, indicated that most participants (86.1\%) rated the hackathon as capable of providing a collaborative and teamwork-oriented environment.

The uncertainty and interdependence features, in turn, influence autonomy, competence, and relatedness \cite{ryan2022self}. \textbf{Autonomy} refers to feeling one has a choice and is willingly endorsing one's behavior \cite{wang2019competence, deci2000and}. In hackathons, autonomy could be reflected in students' ability to self-manage their capability of building a real-world solution in practice, including taking ownership of their learning process \cite{lifshitz2021minimal}. Thus, fostering autonomy in hackathons may promote intrinsic motivation, driving soft skill development as students demonstrate their abilities in a self-directed environment \cite{alkema2017agile}.

In addition, \textbf{competence} refers to the experience of mastery and being effective in one’s activity \cite{wang2019competence, deci2000and}. Hackathons are widely recognized for their ability to foster a multidisciplinary and competitive scenario in which students can showcase their skills \cite{mhlongo2020effectiveness, moshirpour2023multidisciplinary}. As students successfully overcome challenges and contribute meaningfully to their team projects, they experience a sense of competence, reinforcing their learning outcomes. Our case study found that students responded positively to improvements in innovative and creative thinking, collaboration and teamwork, and practical knowledge application, for example.  

Moreover, \textbf{relatedness} refers to the need to feel connected and a sense of belongingness with others \cite{wang2019competence, deci2000and}. In particular, hackathons leverage relatedness by encouraging collaboration, teamwork, and interaction with peers, mentors, and industry professionals \cite{kumalakov2018hackathon}. Through shared experiences, feedback exchanges, and collective problem-solving, students develop a sense of belonging and connectedness, which help to enhance their soft skills \cite{steglich2021online, afshar2022hackathon}.

As previously explained, fulfilling the psychological needs of competence, autonomy, and relatedness affects \textbf{motivation}, which in turn influences the \textbf{outcomes} achieved. Multiple factors could impact students' motivation and, consequently, their learning process \cite{reeve2002self}. In our case study, we noticed different lessons learned. Firstly, the involvement of professors and students in the organizational committee was important, promoting a sense of belonging and ownership among all the stakeholders. Moreover, since this hackathon was the first held on the university campus, most students experienced this type of event for the first time, potentially amplifying their engagement. Engaging alumni from the university enhanced the sense of community and facilitated knowledge sharing. However, alongside these benefits, different challenges emerged. Maintaining student motivation and communication during the hackathon was challenging due to concurrent demands (academic, personal, professional, etc.). Lastly, participants' li\-mited experience with hackathon formats also accentuated the need for eventual introductory workshops to previously cover fundamental topics (\textit{e.g.}, product management or programming), enhancing their readiness for such intensive events.

As a plan for the future, we advocate for understanding psychological needs to drive motivation and enhance students' soft skills through hackathons. Grounded on the discussions derived from the literature, particularly the SDT, and our preliminary results, we may answer our research question by saying that hackathons are able to lead a self-perception of improvement in soft skills for SE students. However, the absence of statistical significance requires careful interpretation and hinders definitive conclusions. We acknowledge that these findings represent an initial stage, but they have also led us to consider another question for the future: \textit{How can hackathons be designed and implemented to better meet SE students' competencies, autonomy, and relatedness, and consequently enhance their soft skills?}

\section{Conclusion}


In this paper, we embrace the idea that hackathons could be a valuable experiential learning strategy for shaping soft skills in SE students in practice. Drawing on Self-Determination Theory and preliminary case study findings, we highlight how the investigated hackathon served as an enabler of an intensive, collaborative environment, where students engaged in software development activities that could be similar to real-world nuances, such as time pressure, collaborative teamwork, innovative thinking, etc. Lastly, future work could investigate how different hackathon formats (e.g., in-person vs. online, short-term vs. extended) potentially affect skill improvement.

{\footnotesize\bibliography{bibliography}}
\bibliographystyle{IEEEtranS}

\end{document}